\documentclass[a4paper, aps, prl, reprint ,superscriptaddress, twocolumn]{revtex4-1}
\usepackage{amsmath}   
\usepackage{graphicx}
\usepackage{hyperref}
\usepackage[utf8]{inputenc}
\usepackage[english]{babel}
\usepackage[sort&compress]{natbib} 

\begin{document}

\title{Role of microscopic phase separation in gelation of aqueous gelatin solutions}

\author{Damjan Pelc}
\affiliation{Department of Physics, Faculty of Science, University of Zagreb, Bijeni\v{c}ka 32, 10000 Zagreb, Croatia}

\author{Sanjin Marion}
\affiliation{Institute of Physics, Bijeni\v{c}ka 46, 10000 Zagreb, Croatia}

\author{Miroslav Po\v{z}ek}
\affiliation{Department of Physics, Faculty of Science, University of Zagreb, Bijeni\v{c}ka 32, 10000 Zagreb, Croatia}

\author{Mario Basleti\'c}
\email{basletic@phy.hr}
\affiliation{Department of Physics, Faculty of Science, University of Zagreb, Bijeni\v{c}ka 32, 10000 Zagreb, Croatia}

\begin{abstract}
Using a unique home-made cell for four-contact impedance spectroscopy of conductive liquid samples, we establish the existence of two low frequency conductivity relaxations in aqueous solutions of gelatin, in both liquid and gel state. A comparison with diffusion measurements using pulsed field gradient NMR shows that the faster relaxation process is due to gelatin macromolecule self-diffusion. This single molecule diffusion is mostly insensitive to the macroscopic state of the sample, implying that the gelation of gelatin is not a percolative phenomenon, but is caused by aggregation of triple helices into a system-spanning fibre network.
\end{abstract}

\date{\today}
\maketitle

Gels are solutions where a small fraction of a gelling solute controls the phase of the whole system\cite{degennes-scaling}. These gelling agents can produce macroscopic disordered networks of interconnected molecules that span the whole solution. Physical properties of gels are mainly governed by the binding energy -- usually of the order of the thermal energy -- and the transition from a fluid (sol) state to a solid-like (gel) state is induced by lowering the temperature. The gelation temperature $T_g$ for this transition can be defined as the temperature where the sample stops to flow (typicaly around $25^\circ$C). This is correlated with the formation of the system-spanning network. 

In this Letter we study a low concentration aqueous solutions of gelatin -- a water soluble polypeptide\cite{Ramachandran1967} made by denaturation from collagen, an irreversible process that destroys one collagen molecule to form three gelatin strands. At low temperatures ($T<T_g$) it forms macroscopically homogeneous and transparent thermoreversible (weak) gels (\textit{gel} state)\cite{Djabourov1988a}.
Above the gelation temperature (at $T_g\sim40^\circ$C), the  solution behaves as a normal viscous liquid (\textit{sol} state). At the molecular level the polypeptide chains have a random coil conformation. By lowering the temperature they slowly change to a single helix conformation. If the concentration of gelatin is sufficient for overlap between nearby molecules, the formation of a triple-helix can be initiated by bonding with up to two other gelatin molecules. The resulting triple-helix is an insoluble semi-rigid rod with the partial structure of collagen due to incomplete renaturation but different energetics\cite{Djabourov1988,Djabourov1988a,Viebke1994,Leikin1994}. This process is the basis for two models of the system-spanning network genesis: (a) bond percolation\cite{degennes-scaling,Djabourov1988} where macroscopic gel forms when the number of intermolecular bonds (triple helices) exceeds a treshold value and a single connected cluster is created; and (b) aggregation\cite{Ren1993,Gornall2007,Parker2010} where triple helices first aggregate into fibres, which then form a macroscopic fibre network. 

Although a widely studied system, there is no consensus on the nature of the sol-gel transition in gelatin. This is unsurprising because gelation is closely related to several other microscopic transitions that span the same temperature ranges like the coil-helix transition and the still not completely understood triple-helix formation. It is known that semi-rigid triple-helices contribute to the elasticity of gels, although it is still unclear if through formation of a percolation cluster or a fibre network\cite{Joly-Duhamel2002}. Recent studies have  shown that there is no simple connection between their macroscopic and microscopic properties\cite{Richter2005,Hagman2010}, making the description of the gelatin gel as a percolation cluster questionable.

Furthermore, the existence of a system-spanning network may impose a restriction on thermal concentration fluctuations, implying an ergodic-nonergodic phase transition (dynamic arrest)\cite{Matsunaga2007,PatrickRoyall2008}, similar to transitions in certain glassy systems\cite{Pusey1989,Ren1993}. In our opinion the nonergodic nature of the transition masks its true behaviour\cite{Pusey1989}, causing inconsistencies in the results of light scattering diffusivity studies in the gel\cite{Amis1983,Ren1992, Blanco2000,Shibayama2001, Okamoto2001, Gupta2005,Matsunaga2007,Pineiro2009}. We propose that combined measurements of frequency dependent conductivity, and diffusivity of gelatin molecules by pulsed-field-gradient nuclear magnetic resonance (PFG-NMR) can resolve those difficulties. In particular, we argue that the system-spanning network is formed by aggregation into fibre networks, as proposed for some biopolymer gels\cite{Viebke1994}. 

The sample used in this work is pig skin gelatin (Sigma-Aldrich, type A, Bloom 300). The solutions were prepared by dissolving gelatin in deionized water and stirring at 60$^\circ$C for 2 hours, erasing any memory the system might have had. Sample volume was typically 
75$\,\mu$l, with gelatin concentrations $c$ in the \textit{dilute} gelling regimes between the minimum gelling concentration $c_\text{gel} \approx 5$ g/l and the overlap concentration $c^*\approx 25$ g/l; and concentrations corresponding to \textit{semi-dilute} unentangled solutions ($c^* < c < c_\text{ent} \approx 150$ g/l)\cite{Guo2003a}.
Gelation is a slow kinetic process, meaning that it is impossible to stop the system at an intermediate step and measure the dynamics. Thus, we did all the measurements after a quench and wait time (non-equilibrium measurements\cite{Blanco2000}), such that we minimize the long length scale structural relaxations but still obtain systematic gelation\cite{Djabourov1988}. Temperature measurements were done while adhering to a strict protocol: a quench ($5\,$K/s) from $50^\circ$C to $5^\circ$C. After waiting for 1 hour for the gelatin to renaturate and relax\cite{Djabourov1988,Gornall2007}, the temperature was raised every $15$ min and after equilibration a spectrum was acquired.  All temperature dependent data are, except when noted differently, given for $10^\circ$C (gel state) and for $42^\circ$C (sol state), well away from the temperature range needed for full denaturation or renaturation of the triple helices\cite{Gornall2007,Parker2010}.
 
\begin{figure}[t!]
\centering
\includegraphics*[height=60mm]{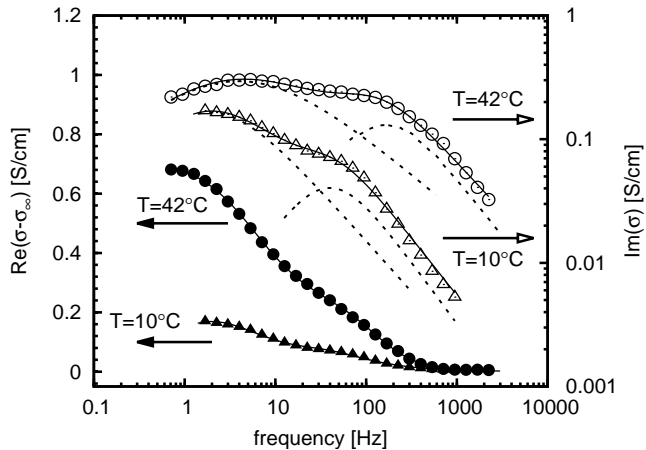}
\caption{Conductivity spectra for a 85 g/l gelatin solution at $42^\circ$C (circles) and $10^\circ$C (triangles). The real (filled symbols) and imaginary (empty symbols) part of the complex conductivity are shown. Solid lines are fits to two Cole-Cole relaxations\cite{Cole1941}. Dotted lines show the individual contributions of the two relaxation processes in $\mathrm{Im}(\sigma)$.  }
\label{fig:impedance_spectrum_50}
\end{figure}

We measure impedance spectra of gelatin solutions in a planar four-contact arrangement, effectively eliminating electrode polarization artefacts\cite{RSI2011}. This technique is similar to dielectric spectroscopy, with the difference that we detect the response of free charges, as opposed to bound charges (e.g.\ dipoles). Thus in our case the appropriate response function is conductivity $\sigma \left( \omega \right)$ instead of permittivity $\epsilon \left( \omega \right)$.
Fig.\ \ref{fig:impedance_spectrum_50} shows (typical) measured spectra for $85\,$g/l gelatin solution, for $T = 10^\circ$C and $42^\circ$C. We note two contributions to the conductivity relaxation, the slow one and fast one. This is in contrast to other impedance spectroscopy results made with two-contact impedance cell (in frequency range from 1 Hz to 10 MHz)\cite{Bohidar1998b, VIEIRA2007}, where low frequency contribution of the impedance is screened by the (unwanted) large electrode polarization effects, typical for conductive liquid samples. Other attempts to mitigate electrode polarization effects had limited success (see references within Ref.\ \cite{RSI2011}), proving that our custom made four-contact setup is an adequate tool for conductivity studies of liquid samples at low frequencies.

Gelatin can be considered a weak polyelectrolyte with a net negative charge of several percent of the number of monomeric units\cite{Rose1987}. Our samples thus consist of free, weakly charged gelatin molecules, dissociated counterions, and any excess salt contained in the sample. Microscopically we can treat the large gelatin molecules as Brownian particles and use the corresponding equation of motion (Langevin equation) to obtain a Drude-like relaxation of the frequency-dependent conductivity,  $\hat{\sigma} (\omega)= \Delta \sigma / (1+ i \omega \tau)$\cite{Kubo-nonequilibrium}. This relaxation has the same form as a Debye relaxation in the dielectric constant. To include the possibility of a spread of characteristic time scales in each process, we analyse our data with the generalized form of the Debye (Drude) formula, the so called Cole-Cole\cite{Cole1941} functional form for complex frequency dependent conductivity, with fast and slow relaxation processes:
\begin{equation}
\hat{\sigma}(\omega)= \sigma_{\infty}+\frac{\Delta\sigma_f}{1+(i \omega \tau_f)^{\beta_f}}+\frac{\Delta\sigma_s}{1+(i\omega\tau_s)^{\beta_s}}
\label{eq:cole-cole}
\end{equation}
where $\tau_{f,s}$ are the characteristic times, $\beta_{f,s}$ exponents corresponding to the spread of relaxation times in the spectrum, $\Delta\sigma_{f,s}$ contributions of the (f)ast/(s)low process to the total conductivity, and $\sigma_{\infty}$ the ``infinite" frequency conductivity due to counterions. Full lines on Fig.\ \ref{fig:impedance_spectrum_50} show Cole-Cole fits to real and imaginary parts of the measured conductivity at fixed temperature, and dotted lines show two individual contributions, `slow'\cite{PripremaSM2012} and `fast', to the $\mathrm{Im}(\sigma)$. From the fits, we can extract characteristic relaxation times $\tau_s \sim 1\,$s and $\tau_f \sim 10\,$ms.

\begin{figure}[t!]
\centering
\includegraphics*[height=60mm]{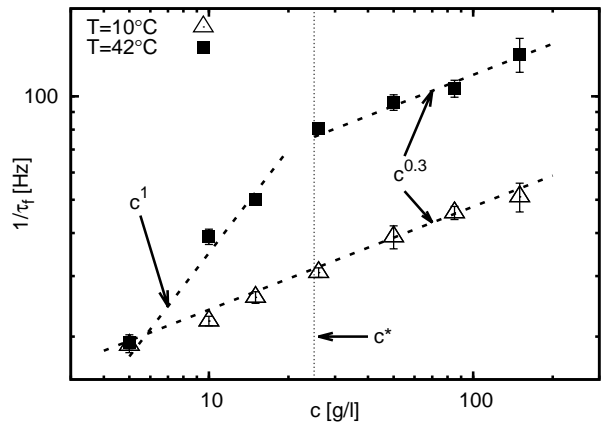}
\caption{ Concentration scaling of the inverse characteristic time for the fast relaxation mode $\tau_f$. The results are given for both the sol and gel state. Concentration $c^*=25$ g/l corresponds to the crossover between dilute and semi-dilute regimes. Inverse characteristic time scales as $1/\tau_{f}\sim c^{0.3}$ in the gel sate and semi-dilute state, while in the dilute sol state it scales as $1/\tau_{f}\sim c^{1}$.
 }
\label{fig:fast:tau_scaling}
\end{figure}

In this work we focus on the faster of the two relaxations ($\tau_f \sim 10$ ms). 
Fig.\ \ref{fig:fast:tau_scaling} shows dependence of the fast process relaxation time, from Cole-Cole fits, on the gelatin concentration $c$. In dilute sol regime ($c<c^*$ and $T=42^\circ$C) relaxation time scales linearly with $c^{-1}$, while in semi-dilute sol regime ($c^*<c<c_\text{ent}$) and all gel regimes ($T=10^\circ$C) we obtain an unusual scaling of $\tau_f \sim c^{-0.3}$. We also note that the relaxation is consistent with a Debye type relaxation, i.e.\ the exponent $\beta_f$ is always equal to $1$ within the experimental error, implying no local heterogeneities for the charge carrier environment. The nature of the relaxation and the obtained exponents indicate that the fast relaxation can be interpreted as a conductivity contribution from gelatin molecule self-diffusion.

To obtain additional information about this self-diffusion process we used pulsed field gradient (PFG) NMR\cite{Walderhaug2010}, with a standard stimulated echo sequence \cite{Hahn1950,Stejskal1965}. The technique enables direct measurement of the dynamic structure factor $S \left(k,t \right)$ by spatially 'tagging' spins using magnetic field gradients (with a linear relationship between wavenumber $k$ and the gradient strength), and recording their position after a time $t$. Experiments were performed on hydrogen nuclei at a NMR frequency of $360$ MHz, in a home-made probe with built-in gradient coils. Dynamic structure factors $S \left(k,t \right)$ obtained from PFG-NMR measurements contain two contributions: a fast 'normal diffusion'  part at times $\sim 10 - 100$ ms over long distances ($\sim 1 - 10$ $\mu$m), consistent with the conventional time and $k$ dependence $S \left(k,t \right) \sim \exp \left( -k^{2} D_{f} t/3 \right)$ and also observed in dynamic light scattering (DLS) studies\cite{Ren1992,Maity1998,Blanco2000}, and an anomalous, slow part at long times ($\sim 1$ s). By comparison of the time-scales involved, we conclude that the two contributions correspond to the two relaxation processes observed in conductivity experiments. The effective diffusion coefficient for the fast process, $D_{f}$, is obtained directly from the short-time slope in a $\ln S$ vs.\ $t$ plot. 
 
\begin{figure}[h!]
\centering
\includegraphics*[width=\columnwidth]{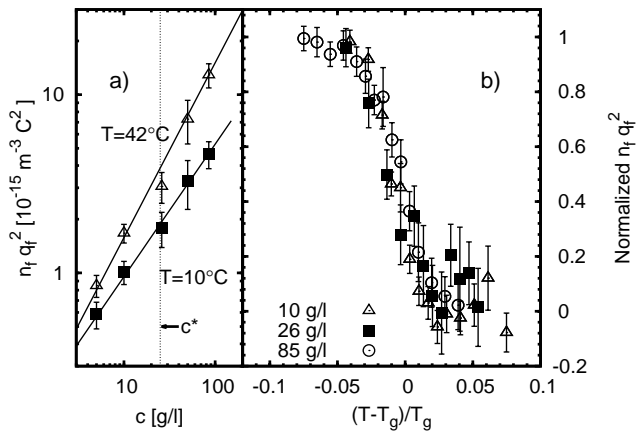}
\caption{ a) The quantity $n_f q_f^2$ obtained from diffusion and conductivity (see text for details). Lines are fits to $ n_f q_f^2  \sim c^{0.98\pm 0.03}$ for the sol, and $n_f q_f^2 \sim c^{0.74\pm 0.02}$ for the gel state.  b) Temperature dependence of the normalized number of gelatin molecules lost to the gel macrostructure ($n_fq_f^2$ data for different concentrations were normalized to the same low and high temperature value).}
\label{fig:fast:nq_scaling}
\end{figure}

By combining the conductivity data and PFG-NMR diffusion we can obtain a relation between the concentration of charge carriers and dissolved gelatin. The Langevin equation approach provides us with a simple (Einstein) equation which connects the characteristic relaxation height $\Delta\sigma_f$ and the diffusion constant $D_f$ with the effective number of charge carriers, $n_f$: $\Delta \sigma_f = n_f q_f^2 D_f / kT$\cite{Kubo-nonequilibrium} (with $q_f$ the effective charge of the carriers). We can now use the concentration dependent values of $\Delta\sigma_f$ and $D_f$ to obtain the concentration scaling  of $n_f q_f^2$. This is shown in {Fig.\ \ref{fig:fast:nq_scaling}a where a scaling of the form $ n_f q_f^2 \sim c^b$ is obtained with $b=(0.98 \pm 0.03)$ in the sol and $b=(0.74 \pm0.02)$ in the gel state. This scaling is unchanged at the transition concentration $c^*$ between the dilute and semi-dilute regime, proving that we are indeed dealing with self-diffusion of gelatin molecules. The sol state scaling shows that there is a linear relationship between the number of charge carriers and concentration of gelatin, while in the gel state we lose a concentration dependent fraction of charge carriers due to gelation (see Fig.\ \ref{fig:fast:nq_scaling}a). The reason for the incomplete renaturation most likely lies in the inherent randomness between gelatin molecules at the molecular level\cite{Djabourov1988a}. Due to a few orders of magnitude faster renaturation\cite{Kubelka2004} and the limited contribution of helix formation to rheology\cite{Gornall2007} we find that the $n_fq_f^2\sim c^{-0.74}$  scaling in the gel state is consistent with the picture that a fraction of the molecules at any time are a part of the gel macrostructure and thus cannot contribute to the conductivity. Fig.\ \ref{fig:fast:nq_scaling}b shows the reduced temperature dependence of $n_f q_f^2$ normalized to their high temperature value, for several concentrations. All data conform to a master curve, with the point of inflection roughly corresponding to the gelation temperature. This is consistent with light polarization studies where the same behaviour is seen on the level of triple-helix formation (triple-helix melting curve)\cite{Djabourov1988,Guo2003,Guo2003a,Gornall2007}. Preliminary circular dichroism (CD) results\cite{PripremaSM2012} also indicate that a master curve exists for the concentration of triple-helices and that it has the same form, although the transition temperature obtained from conductivity is somewhat higher and with a different concentration dependence. The numerical values for $n_f q_f^2$ for the case of $\sim 100$ monomeric units in our gelatin indicate that about 2-3\% of the gelatin monomers have dissociated, as would be expected for a weak polyelectrolyte.

\begin{figure}[h!]
\centering
\includegraphics*[height=60mm]{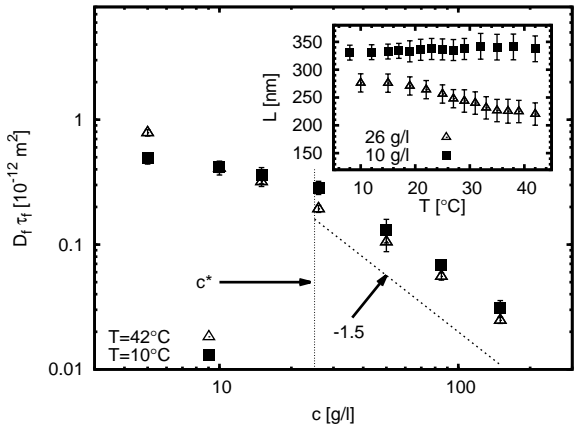}
\caption{ Concentration scaling of the characteristic length scale for self-diffusion $L^2= D_f \tau_f$. For concentrations larger than the overlap concentration $c^*=25$ g/l the data is consistent with a scaling of the form $L^2 \sim c^{-1.5}$  for both sol and gel. Below $c^*$ the exponent reduces to $L^2 \sim c^{-0.85}$ for the sol state, while for the gel state scaling exponent is not well defined. Inset shows the characteristic length scale $L$ as a function of temperature $T$.}
\label{fig:fast:L_scaling}
\end{figure}

We can also obtain a characteristic length scale $L$ associated with the self-diffusion process by using the known scaling relation $L^2\sim D_f \tau_f$, as shown in Fig.\ \ref{fig:fast:L_scaling}. The weak temperature dependence of $L$ for two concentrations shown in the inset of Fig.\ \ref{fig:fast:L_scaling}  shows that $L$ is a well defined length scale, roughly independent of temperature and thus of the macroscopic state in the entire concentration range. However, the dynamics leading to a well-defined $L$ is markedly different for dilute and semi-dilute solutions. The behaviour of dilute solutions in the sol state ($c_g<c<c^*$, $T=42^\circ$C) is consistent with a picture of diffusing independent swollen coils: the scaling $\tau_f\sim c^{-1}$ (Fig.\ \ref{fig:fast:tau_scaling}) shows that the time between collisions is proportional to the available volume per molecule, and for the characteristic length we obtain $L^2 \sim c^{-0.85}$, close to the scaling expected for non-overlapping spheres $L^2\sim c^{-2/3}$. The value of the diffusion constant indicates a gyration radius for the swollen coils of $R_g\approx 20$ nm. Scaling for the dilute gel state ($c_g<c<c^*$, $T=10^\circ$C) turns out not to be well defined, which we speculate is due to the complexity of the gel forming below the polymer overlap concentration, with possible competition between aggregation and percolation.

In the semi-dilute sol and semi-dilute gel states ($c>c^*$) we believe we are detecting translational diffusion resulting from reptation of gelatin macromolecules. We find $L^2\sim c^{-1.5}$, consistent with the scaling of the mesh size from the De Gennes theory of polymers in good solvents\cite{degennes-scaling}, with the mesh size $\sim 100 - 300$ nm (slightly larger than indicated previously\cite{Ren1992,Blanco2000,Barrera2010}). DLS studies report an analogous mode having a scaling in semi-dilute $D\sim c^{-1.75}$\cite{Amis1983, Ren1992}, while our results indicate a lower value $D\sim c^{-1.2}$. Although in the case of reptation one would expect a non-Debye relaxation and  anomalous diffusion in PFG-NMR, we emphasize that we are working at longer length-scales than DLS: at our length scales ($\sim \mu$m) the reptation dynamics renormalizes into the translational diffusion $D_f$ of the center of mass of the gelatin molecule\cite{Doi1986}.} This makes PFG NMR better suited for comparison with conductivity studies. Both the theory and our experimental data show that the corresponding characteristic length $L$ scales as the mesh size for (non-charged) semi-dilute polymers. If we now consider the gel state in the semi-dilute regime, we have the same length scaling and time dependence as in the sol. We thus conclude that the self-diffusion must be independent of the macroscopic state of the sample, depending only on the existence of a polypeptide mesh, originating from either chain overlap or gelation. This confirms that rheological properties and diffusion of gelatin are independent; there are no heterogeneities in the local environment of mobile gelatin molecules. We also observe that the length scale concentration dependence defining a mesh is almost unchanged due the sol-gel transition.

The concentration of non-bound gelatin as implied by Fig.\ \ref{fig:fast:nq_scaling}a and \ref{fig:fast:L_scaling} is sufficient to initiate the formation of triple-helices, but no such effect is seen in self-diffusion. This shows that after gelation we have two phases with a fast exchange of gelatin molecules, one analogous to the sol phase but less concentrated, and an aggregated phase of renaturated gelatin forming a fibre network. Gelatin, like collagen and similar systems\cite{Tempel1996,Lieleg2011}, can form fibres\cite{Djabourov1988} due to the attractive interaction between triple-helices in collagen\cite{Kornyshev2007}. These interactions lead to spinodal decomposition in a bad solvent (e.g.\ water-methanol), but in pure aqueous dilute and semi-dilute solutions, like the one studied here, there is no competition between gelation and precipitation because of the quality of solvent\cite{Tan1983,Bohidar1993a}. Thus the aggregation we observe can be regarded as a form of microphase separation distinct from spinodal decomposition\cite{degennes-scaling}. The network of aggregated fibres is responsible for macroscopic properties like elasticity, explaining why no simple correlation exists between gel strength (macrostructure) and diffusivity of probe particles\cite{Hagman2010} and why microscopic probes like light scattering predict a different gelation transition point from macroscopic ones like rheology\cite{Richter2005}. 

Because the gelation process was previously connected with percolation, current theories for similar systems do not account for the phase separation in the good solvent regime\cite{Semenov1998,Rubinstein1998} due to neglected microscopic cooperativity. Available theories for thermoreversible gelation (see references in \cite{Rubinstein1999} and \cite{Tanaka2002,Tanaka2004,Tanaka2006}) and numerical studies\cite{Suarez2009a} do not predict the concentration scaling for the diffusion and relaxation times nor the gelling behaviour in the dilute regime. Our results lead to the conclusion that gelatin is closer to systems showing phase separation with gelation like colloid gels\cite{Ronsin2009} and similar systems\cite{Tanaka2005,Lu2008}. All this would make gelatin an excellent system for studying the complex phenomena of fibre networks\cite{Picu2011,Lieleg2011,Broedersz2011a}, e.g. glassy behaviour under stress\cite{Parker2010}.

In summary, by applying a four-electrode impedance spectroscopy method to dilute and semi-dilute gelling gelatin solutions we show the existence of two processes that contribute to the frequency dependent conductivity. We identify the fast process as gelatin molecule self-diffusion due to long length scale reptation dynamics (translational diffusion). Self-diffusion at the crossover from dilute to semi-dilute state indicates that the microscopic diffusivity of particles is decoupled from the macroscopic properties of the gel. Due to a loss of charge carriers in the gel we argue that upon gelation there exist two microphases in the system: the sol microphase where there is no interconnectivity and no difference from the sol macrophase, and the fibre microphase formed via aggregation of triple-helices into fibre networks. Our results  place semi-dilute (and dilute) gelatin solutions away from the standard bond percolation model, and they suggest that the gelation process goes via aggregation into fibre networks. This, although implied by the similarity of rheological data between known fibre-formers (e.g. actin)\cite{Storm2005}, has not yet been established. Thus, studies of gelatin networks could give additional insight of the entire class of fibre forming systems.

\begin{acknowledgments}
We are pleased to acknowledge the assistance of Nata\v{s}a \v{S}ijakovi\'c-Vuji\v{c}i\'c and helpful discussions with Tomislav Vuleti\'c. The research leading to these results was supported by equipment financed from the European Community’s Seventh Framework Programme (FP7/2007-2013) under
grant agreement no.\ 229390 SOLeNeMaR, and by funding from the Croatian Ministry of Science, Education and Sports through grants 
no.\ 119-1191458-1023 and 119-1191458-1022.
\end{acknowledgments}

\end{document}